# Devices for Thermal Conductivity Measurements of Electroplated Bi for X-ray TES Absorbers

Orlando Quaranta[1] • Lisa M. Gades[1] • Cindy Xue[1] • Ralu Divan[1] • Umeshkumar M. Patel[1] • Tejas Guruswamy[1] • Antonino Miceli[1]

[1]*Argonne National Laboratory*

**Abstract** Electroplated Bismuth (Bi) is commonly used in Transition-Edge Sensors (TESs) for X-rays because of its high stopping power and low heat capacity [1, 2]. Electroplated Bi is usually grown on top of another metal that acts as seed layer, typically gold (Au), making it challenging to extrapolate its thermoelectric properties. In this work, we present four-wire resistance measurement structures that allow us to measure resistance as a function of temperature of electroplated Bi independently of Au. The results show that the thermal conductivity of the Bi at 3 K is high enough to guarantee the correct thermalization of X-ray photons when used as an absorber for TESs.

## 1 Introduction

Transition-Edge Sensors (TESs) are seeing increasing use in synchrotron X-ray facilities. Their energy resolution, 10-100x better than comparable silicon-drift diode sensors, enables new science in the areas of X-ray emission spectroscopy [3], Compton scattering [4], and XAFS [5]. One of the challenges in developing X-ray TESs is to fabricate a suitable absorber with sufficient X-ray stopping power (energies in the 6-100 keV range) while also retaining good thermalization properties to avoid position-dependent responses and keeping the total heat capacity limited to retain high energy resolution. Gold (Au) and Bismuth (Bi) form a common combination that works because both have good stopping power, Au is characterized by very high thermal conductivity [6] even at cryogenic temperatures, and Bi has lower heat capacity [1] due to its limited number of carriers. At the same time, this could affect the thermal conductivity of the material at cold, negatively affecting the performance of the sensor.

The resistivity ($\rho$) is a common proxy for describing thermal properties of materials at lower temperatures. Previous work has estimated a lower $\rho$ for

## Orlando Quaranta • Lisa M. Gades • Cindy Xue

electroplated Bi compared to thermally evaporated Bi [7]. However, the presence of an electrically conductive Au seed layer underneath an electroplated Bi film poses a difficulty for directly measuring $\rho$. In this paper, we describe the deployment of a four-wire resistance measurement structure fabricated to isolate the Bi contribution and thus allow a more direct measurement of its electrothermal properties at cryogenic temperatures.

## 2 Device design and fabrication

2.1 Device design

The electroplating of Bi is usually obtained by electrochemically growing the Bi on top of a seed layer. This is convenient because for X-ray TESs, Bi is always coupled with another absorbing material, typically Au. If the four-wire measurement structure was to be fabricated directly on the stack of Bi and Au, $\rho$ measurements would be dominated by the higher conductivity of the Au layer [6]. Some precautions can be taken to minimize the effect, like using thinner Au layers and tuning the Bi electroplating conditions to maximize its conductivity [8], but ultimately the convolution of the two contributions will be measured.

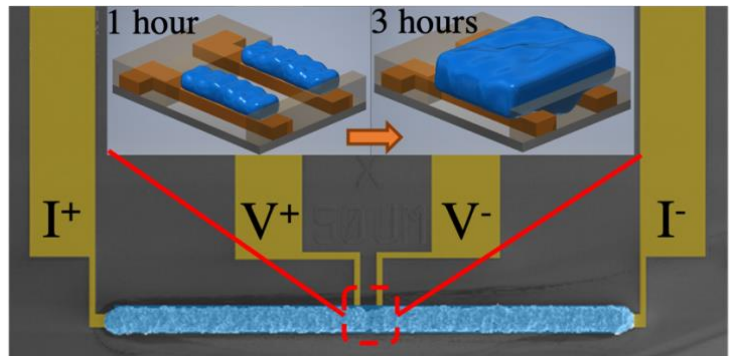

**Fig. 1** SEM image in enhanced colors showing the devices and labels for the four-wire measurements leads. The Au bond pads and leads are represented in yellow, while the Bi micro-wire is represented in blue. In gray is represented the SiO$_2$ substrate. In the insert is a 3D rendition of Bi resistivity test structures zoomed in around the voltage lead area. In the insert, in orange is represented the Au layer, in blue is the Bi, in gray is the SiO$_2$ substrate and in transparent is the photoresist mold. The figures represent the Bi growth after 1 hour (*Left*) and 3 hours (*Right*) of electroplating respectively. After 3 hours, the Bi micro-wire is completely formed. (Color figure online)

# Devices for Thermal Conductivity Measurements of Electroplated Bi for X-ray TES Absorbers

To measure the electroplated Bi resistivity independently from the underlaying Au layer, the devices represented in the SEM image in Fig.1 have been fabricated. In the insert of Fig. 1 is a 3D rendering of the same device zoomed in on the area between the voltage leads. The devices are composed of four separate leads patterned in a Au thin film, which also acts as a seed layer for the electrochemical growth of the Bi. During the electrochemical deposition the Bi grows initially on top of the Au seed layer. The growth is somewhat isotropic, unless otherwise constricted. In these devices, a thick photoresist mold is present, which directs the growth of the Bi mainly vertically out of the plane of the wafer and subsequently parallel to the wafer plane, in the direction that connects the four leads. When enough material has been deposited, the Bi grown from each lead eventually meets in the middle of the mold, forming a continuous microwire of Bi, which represents the only element of electrical continuity between the Au leads. Therefore, the voltage drop between the two inner leads is determined only by the Bi resistance. Structures of several dimensions have been fabricated to study the limits of this approach and to check that the measured resistances would scale with the device lateral dimensions. A length of 100 μm between the voltage leads represented the maximum that could be achieved reliably. Devices of three different widths have been fabricated at this length: 20, 50 and 100 μm. All the devices were fabricated on a $SiO_2$/Si wafer to minimize current leakage through the wafer at all temperatures.

2.2 Device fabrication

Au was used as the conductive seed layer for plating Bi and also formed the measurement pads for the four-wire resistance measurement structures. 100 nm Au with a 5 nm Ti adhesion layer was deposited by sputtering and patterned by liftoff (lithography via Heidelberg MLA150). A ~23 μm layer of AZ P4620 photoresist formed the Bi plating mold. This required a two-layer spin process; longer, cooler soft-bake cycles; and multiple exposures in the MLA150 to prevent bubbling in the photoresist.

The wafer was prepared for plating with a 1-min O2 plasma RIE descum and a 10-sec dip in dilute nitric acid for surface wetting. The electrolyte consisted of bismuth-nitrate pentahydrate dissolved in a mixture of nitric acid, glycerol, and tartaric acid, buffered with KOH to a pH of ~0.15 (process was adapted from [9]). For this electrolyte, maintaining a pH below ~0.3 is required to prevent Bi precipitates. However, plating is inhibited if pH falls close to 0. The electrolyte pH may be adjusted simply by adding either nitric acid or KOH, as required. Plating was completed in the beaker-suspended version of the "Beaker on a Stick" wafer plating holder from Wafer

# Orlando Quaranta • Lisa M. Gades • Cindy Xue

Power Technologies [10]. Wafers were plated at room temperature, and the solution was not agitated [8]. Plating required a low-throw process; the current density for these wafers was 7.7 mA/cm$^2$. A given wafer was plated for 60 min, at which time the Bi thickness was estimated via profilometry, measuring the difference between the photoresist mold of known thickness and the surface of the Bi film. The plating rate was estimated to be ~220-250 nm/min in the larger Bi structures. The wafer was then plated an additional 90-120 min to allow enough lateral growth for the Bi to form a bridge between the Au fingers.

The wafer was rinsed in dilute nitric acid to prevent Bi precipitates on the plated surface, followed by DI water. A protective layer of AZ P4620 covered the Bi structures for dicing. Chips were diced, and photoresist was removed by acetone and isopropyl alcohol before measurement.

## 3 Results and discussion

The resistance ($R$) of Bi microwires of various dimensions has been measured as a function of temperature ($T$) from 300 K to 3 K. In Tab. 1 are reported the measured resistance values at 300 K for devices of the same length ($l$) but of different width ($w$) averaged over several samples. The devices all have a thickness of approximately 30 μm. For reference, the expected resistance for each type of device, based on the resistivity for bulk Bi, is also reported [6].

| $l \times w$ (μm x μm) | Bulk R @ 300 K (Ω) | Device R @ 300 K (Ω) | RRR | $\kappa$ @ 3 K (W*m$^{-1}$*K$^{-1}$) |
|---|---|---|---|---|
| 100 x 20 | 0.198 | 0.190 | 0.41 | 4.62 x 10$^{-2}$ |
| 100 x 50 | 0.086 | 0.086 | 0.35 | 3.47 x 10$^{-2}$ |
| 100 x 100 | 0.041 | 0.042 | 0.32 | 3.44 x 10$^{-2}$ |

**Tab. 1** Averaged measured resistance at 300 K, RRR and thermal conductivity at 3 K for various test structures. Expected resistance at 300 K from bulk resistivity [6] is also reported. All devices are approximately 30 μm thick ($t$).

The averaged measured resistance at 300 K closely matches the expected resistance from bulk resistivity [6]. The devices whose resistance differ the most from the expected value are the 100 x 20 variant. This is probably due to the difficulty in properly defining the actual device dimensions due to the large grain size of Bi in the micro-wire, grains being several μm wide. The close match between expected and measured resistance provides confidence in the validity of the structure for the measurement of electroplated Bi resistance.

# Devices for Thermal Conductivity Measurements of Electroplated Bi for X-ray TES Absorbers

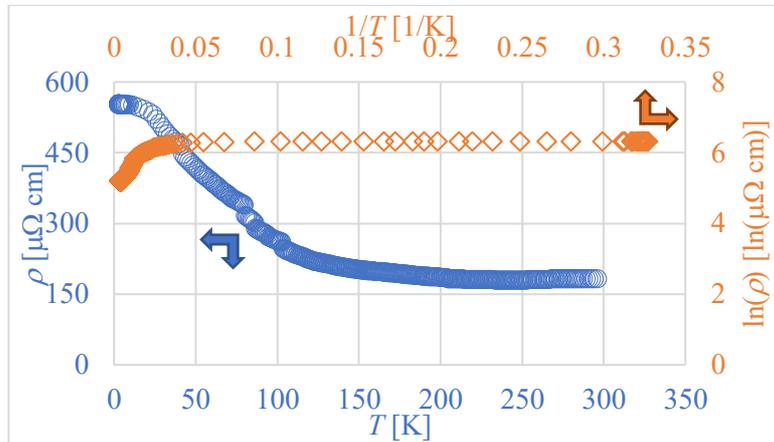

**Fig. 2** Resistance versus temperature curve in *blue open circles* on *Left* and *Bottom axes*. Natural logarithm of the resistivity in *orange open diamonds* on *Right* and *Top axes*. All measurements are from a representative device. (Color figure online)

Shown in Fig. 2 is an example of a $\rho$ vs $T$ curve for one of the devices (blue open circles on bottom and left axes) measured from 300 K down to 0.1 K. The resistivity initially falls approximately linearly with temperature down until about 230 K. Below this temperature, the resistivity starts to rise all the way down to approximately 10 K, at which point it saturates. At first impression, the rise in resistivity resembles that of a semiconductor, but this isn't quite correct. The first sign against the semiconductive nature of the microwire is the fact that the resistivity saturates a low temperature, instead of diverging as it should in the presence of an energy gap in the carrier density of states. One possible explanation for the saturation effect is the activation of some form of extra conduction channels at lower temperatures, which has been hypothesized in Bi crystals [11]. The second sign against the semiconductive nature of the device is the non-exponential rise in resistance. The orange diamonds in Fig. 2 show the natural logarithm of the resistivity as a function of the inverse of the temperature (top and right axes); the curve doesn't show any obvious linear dependence. Conversely, a non-exponential rise of the resistivity with lowering temperatures can be explained in terms of the temperature dependence of the overlapping valence and conduction bands in a semimetal [11]. Similar trends in the resistance versus temperature have been seen in Bi crystals under pressures [11]. Our hypothesis is that when the crystals that form the microwire grow from two opposite directions from the voltage leads and connect in the middle, they push against each other. When

**Orlando Quaranta • Lisa M. Gades • Cindy Xue**

the microwire is cooled down, the mismatch between the thermal contraction of the Si substrate and that of the Bi crystals causes the formation of mechanical stress, similar to that reported in [11]. Further studies are needed to completely understand this phenomenon.

Knowledge of the low temperature resistivity of the microwire allows us to estimate the thermal conductivity at cold. The thermal conductivity ($\kappa$) in metals is proportional to the electrical conductivity ($\sigma$) through the Lorenz number ($L$), as stated by the phenomenological law of Wiedemann-Franz

$$\kappa = \sigma LT. \tag{1}$$

The Lorenz number is a material-dependent number. We derived an estimate for our devices by using $\kappa$ and $\rho$ for bulk Bi at 300 K from literature [6]. The approach is justified by the good agreement between the measured electrical resistivity and the resistivity for bulk, as shown in Tab. 1. The resulting $\kappa$ at base temperature is reported in Tab. 1.

The Bi microwire devices have an average $\kappa$ about 2 orders of magnitude lower than that of the Au film used as seed layer and absorbers in our sensors (~7.8 Wm$^{-1}$K$^{-1}$). This is expected given the semimetal nature of Bi. At the same time, the thermal conductivity of the SiN membranes in our devices is about 3 orders of magnitude lower than that of Bi (~1.2x10$^{-5}$ Wm$^{-1}$K$^{-1}$), estimated from the current-voltage characteristics of our sensors. This guarantees that the heat generated by the absorption of an X-ray photon in the Au/Bi absorber stack will thermalize within the TESs before escaping to the thermal bath through the SiN membrane.

## 4 Conclusions

In this work, we described the fabrication of test structures for measuring thermoelectric properties of electroplated materials while eliminating the contribution from the undelaying seed layer. We showed that the electroplated Bi typically used as an X-ray absorber in TESs has an acceptable thermal conductance at operational temperatures, guaranteeing that even at the high thicknesses needed for absorption of high-energy photons, no position-dependence effect should take place.

Future works will focus on studying the possibility of the microwire going through a semimetal-to-semiconductor transition when cooled below ~ 250 K. This could be the effect of the mechanical stress induced in the Bi grains by differential contraction between the substrate and the microwire, but more studies are needed.

# Devices for Thermal Conductivity Measurements of Electroplated Bi for X-ray TES Absorbers

**Acknowledgements**

This research is funded by Argonne National Laboratory LDRD proposals 2018-002-N0 and 2021-0059; is supported by the Accelerator and Detector R&D program in Basic Energy Sciences' Scientific User Facilities (SUF) Division at the Department of Energy; uses resources of the Advanced Photon Source and Center for Nanoscale Materials, U.S. Department of Energy (DOE) Office of Science User Facilities operated for the DOE Office of Science by the Argonne National Laboratory under Contract No. DE-AC02-06CH11357; and makes use of the Pritzker Nanofabrication Facility of the Institute for Molecular Engineering at the University of Chicago, which receives support from Soft and Hybrid Nanotechnology Experimental (SHyNE) Resource (NSF ECCS-1542205), a node of the National Science Foundation's National Nanotechnology Coordinated Infrastructure.